\documentclass[aps,prl,twocolumn,showpacs]{revtex4}
\usepackage{epsfig}

\usepackage{color}

\newcommand{\be}{\begin{equation}}
\newcommand{\ee}{\end{equation}}
\newcommand{\bea}{\begin{eqnarray}}
\newcommand{\eea}{\end{eqnarray}}

\begin{document}

\title{Color Superfluidity and ``Baryon'' Formation in Ultracold Fermions}
\author{\'Akos  Rapp$^{1,2}$, Gergely Zar\'and$^{1,2}$, Carsten Honerkamp$^3$, and  Walter Hofstetter$^4$}
\affiliation{
$^1$ Theoretical Physics Department, Institute of Physics, Budapest University of Technology and Economy,
Budapest, H-1521, Hungary,\\
$^2$ Institut f\"ur Theoretische Festk\"orperphysik,Universit\"at Karlsruhe, D-76128 Karlsruhe, Germany,\\
$^3$Theoretical Physics, Universit\"at W\"urzburg, D-97074 W\"urzburg, Germany\\
$^4$ Institut f\"ur Theoretische Physik, Johann Wolfgang Goethe-Universit\"at, 60438 Frankfurt/Main, Germany 
}

\date{\today}

\begin{abstract}
We study fermionic atoms
of three different internal quantum states (colors) in an optical lattice, which are 
interacting through attractive on site interactions, $U<0$. 
Using a variational calculation for equal color densities  and small couplings,  $|U|  < |U_C|$, 
a color superfluid state emerges with a tendency to  domain formation.   For $|U|  > |U_C|$, triplets of atoms with different colors form singlet fermions (trions). 
These phases are the analogies of the color superconducting and baryonic phases in QCD. 
In ultracold fermions,  this transition is found to be of \emph{ second order}. 
Our results demonstrate that quantum simulations with ultracold gases may shed light 
on outstanding problems in quantum field theory. 
\end{abstract}

\pacs{03.75.Mn , 32.80.Pj, 71.35.Lk}
\vspace*{-20pt}
\maketitle

The achievement of Bose-Einstein condensation (BEC) a decade ago has  
opened the new field of ultracold atomic physics where dilute alkali-metal gases are cooled  
into the quantum degenerate regime \cite{BEC}. 
In particular, degenerate Fermi gases have been realized \cite{fermions} with temperatures down to 
$T/T_F \approx 0.05$. An attractive feature of these systems is the high degree of 
\emph{ tunability}. Feshbach resonances allow atomic interaction strengths to be tuned \cite{feshbach}, while 
optical lattices can be used to create artificial crystals of light in which atoms form 
Bloch bands like electrons in solids \cite{lattice-1d,Mott,lattice-theo}. 
Moreover, the effective interaction 
strength can be modulated by the optical lattice alone and thus tuned through quantum phase 
transitions like the bosonic Mott transition \cite{Mott}. 
Recently, fermionic $^{40}$K atoms have been loaded into optical lattices as well \cite{lattice-fermions}. 
It has been predicted that quantum simulations with cold fermions in optical lattices 
may ultimately shed light on complex solid-state phenomena like high-temperature 
superconductivity \cite{HTC}. 
In a different line of experiments, fermionic superfluids close to the BEC-BCS transition  
have been created and even loaded into optical lattices \cite{bec-bcs,bec-bcs-lattice}. 
Very recently, also pairing of fermions with unequal spin densities and the eventual breakdown of 
superfluidity has been studied \cite{bcs-imbalance}. 

More generally, with the degrees of freedom offered by cold atoms, it is possible to create 
new states of matter which have no equivalent in condensed matter. 
While typical electronic systems have at most SU(2) spin rotational symmetry, 
the atomic total angular momentum $F$ can be larger than $1/2$, resulting in $2F+1$ hyperfine states. 
In optical traps, all of these states can be trapped, e.g., for spinor condensates 
of $^{23}$Na and $^{87}$Rb  \cite{spinor}. 
For fermionic atoms, coexistence of 3 different hyperfine states of $^{40}$K in an optical trap 
has been demonstrated \cite{3states}. 

Here, we study optical lattices  loaded with fermionic atoms that possess 
three different internal quantum numbers (colors), $\alpha = 1\dots 3$ and interact through 
an attractive on-site interaction. This system is well described by a SU(3) Hubbard Hamiltonian.
For small interactions a color superfluid (CSF) state appears with a triplet order parameter~\cite{Honerkamp1}.
 Here  we show that for low fillings
and larger interactions a quantum phase transition takes place from the superfluid state to a 
Fermi liquid phase, where groups of three fermions bind together to form \emph{ trions}. 
This transition is closely analogous to the one conjectured in QCD, 
where at large quark densities, a color superconducting state is expected while at 
low densities, baryonic matter emerges. 

Such an optical  system can be realized, e.g., by loading 
$^6$Li atoms of nuclear spin $I=1$ into an optical lattice and applying a magnetic field  
that is larger than the hyperfine coupling. In this case, the electron spin of the Li atom is essentially 
polarized along the external field direction and the $2I+1$ nuclear quantum numbers provide 
the internal degrees of freedom. Attractive interactions independent of the nuclear spin 
are induced by an anomalously large and negative triplet scattering length $a_s = -2160 a_0$ \cite{Li6}.
We expect the experimentally accessible temperatures for this three-state ensemble 
to be in the same range as for two-component mixtures, i.e., $T/T_F \approx 0.05$ \cite{bec-bcs}. 
In the absence of spin-flip processes, the densities of the three ``color'' states can be adjusted 
independently by radiofrequency sweeps and selective evaporation \cite{bcs-imbalance}. 
 
In an optical lattice the atoms experience a periodic potential $V(x) = V_0
\sum_{l=x,y,z} \cos^2 (k x_l)$, with    
$k$ the wave vector of the laser and the directions labeled by $l$. 
At low fillings, this system can be described with good accuracy by the  
Hubbard Hamiltonian
\be
H= - t {\sum_{i,j,\alpha}} \hat c_{i\alpha}^\dagger \hat c_{j\alpha} 
+ \frac{U}{2} \sum_{i,\alpha\ne\beta}  \hat n_{i\alpha} \hat n_{i\beta}\;.
\label{eq:H}
\ee
Here  ${\hat n}_{i\alpha} =  \hat c_{i\alpha}^\dagger \hat c_{i\alpha} $ measures the number of fermions 
with color $\alpha$ at site $i$,   with $\hat c_{i\alpha}^\dagger$ the creation operator of a fermion.
The interaction $U$ is assumed to be negative throughout this Letter while the ratio 
$|U|/t$ can be tuned in a wide range by changing the depth $V_0$ of the optical lattice. 
In Eq.~(\ref{eq:H}), we also assume a homogeneous optical lattice, and neglect to a first approximation 
the parabolic confinement potential of the atom trap. 
Note that Eq.~(\ref{eq:H}) has an SU(3) symmetry associated with global rotations, $c_{i\alpha} \to\sum_\beta U_{\alpha\beta}
c_{i\beta}$, and $H$ conserves the total number of particles of 
each  color, ${\hat N}_\alpha \equiv \sum_i {\hat n}_{i\alpha}$.  
For $^6$Li in a strong magnetic field,
this symmetry arises because the interaction between the atoms is mediated by electronic van der Waals
forces, which are independent of the nuclear degrees of freedom.  

For small values of $|U|/t$, the attractive interaction has been 
shown to induce a  superfluid state at $T=0$ \cite{Honerkamp1}. 
In this state, fermions are paired to form Cooper pairs with an order parameter 
$\Delta_{\alpha\beta} \equiv \langle c_{i\alpha}c_{i\beta}\rangle$
which transforms according to the adjoint representation $\bar{3}$ of SU(3).
For very large  values of $|U|/t$, however, perturbation theory in the hopping $t$ 
predicts a state of entirely different nature: Here  three atoms tend to form a composite object, 
a local bound color singlet of energy $3U$, which can hop between neighboring sites 
with an amplitude $t_{\rm eff} \sim {t}^3/U^2$ and interact repulsively to form a Fermi liquid at very low 
temperatures. Note that the symmetry of the superfluid and the trionic
state is   different;  therefore, there must be a phase transition that separates them.
As we shall see, this transition is of second order, and the two states are separated by a 
quantum critical point. 

The above transition occurs at intermediate values of $U/t$, where 
neither perturbation theory nor diagrammatic approaches can be applied successfully. 
We therefore use a variational approach in the high-dimensional limit $d\to\infty$. 
As a variational ansatz at $T=0$, we use a Gutzwiller-projected wave function 
$
\vert G \rangle = \prod_i g^{\hat t_i} \vert BCS \rangle \;, 
$
where $\vert BCS \rangle$ is a BCS superfluid state with two out of three 
color states paired 
\be
| BCS \rangle= \prod_{\epsilon_{{\bf k}'} <\mu_3} \hat c_{{\bf k}' 3}^+ \prod_{{\bf k}} 
(u_{{\bf k}}+v_{{\bf k}} \hat c_{{\bf k} 1}^+ \hat c_{-{\bf k} 2}^+)|0\rangle \;,
\label{eq:BCS}
\ee
and $\hat{P}_G\equiv  \prod_i g^{\hat t_i} = \prod_i [1 + (g-1) \hat n_{i1} \hat n_{i2} \hat n_{i3}]$ is a generalized Gutzwiller projector that favors triply occupied states for large values of the variational parameter $g$.
Without making a restriction,  we have assumed in Eq.~(\ref{eq:BCS}) that only the $\Delta_{12} = -\Delta_{21}$ components of the order parameter are non-zero.This corresponds to a particular gauge of the pairing field \cite{Honerkamp1}. 
The factors  $u_{\bf k}$ and $v_{\bf k}$ denote the usual BCS  coherence factors, 
$u_{\bf k}^2 = \frac{1}{2} \bigl(1+ {\xi_{\bf k}}/{\sqrt{\xi_{\bf k}^2 + \Delta^2}} \bigr)$
and $v_{\bf k} = \sqrt{1- u_{\bf k}^2}$, with  
$\xi_{\bf k} =\epsilon_{\bf k} -\mu_{12}$, and 
 $\epsilon_{\bf k} =  -2 t  \sum_{l=1}^d \cos k_l $ 
the kinetic energy of the atoms on the lattice.

For  $n_1=n_2=n_3$, one can gain energy  in the superfluid state 
by transferring particles to the channels where the superfluid  condensate forms. 
Therefore, for a given total density $n$, the CSF has  
a global energy minimum for slightly unequal densities, $n_1=n_2\ne n_3$. 
In Eq.~(\ref{eq:BCS}), we therefore 
introduced different chemical potentials for the first two and the third channel,  $\mu_{12} \ne \mu_{3}$.
Note, however, that  $\mu_{12} \ne \mu_{3}$  are not our variational parameters;
they are just used to fix the total density $n \equiv \sum_{\alpha=1}^3 \langle {\hat n}_{i\alpha} \rangle$ and  
$n_3 \equiv   \langle {\hat n}_{i3}\rangle$. 

To obtain our variational estimate for the energy, we need to evaluate 
the expectation value of the Hamiltonian,  $E(\Delta,g,n,n_3) \equiv \langle G| H | G \rangle/\langle G|G\rangle$, 
and minimize it with respect to the independent parameters $\Delta$, 
$g$ and $n_3$ for a given density $n$. 
To compute quantities of the form 
$\langle \hat O\rangle_G\equiv \langle BCS| {\hat P}_G \hat O {\hat P}_G | BCS\rangle$, 
we first  expand the product  
$\hat P_G = \prod_i [1 + (g-1) {\hat t_i}] $
and  rewrite ${\hat P}_G \hat O {\hat P}_G $ so
that no operator $\hat c^\dagger_{i\alpha}$ occurs to the right of the corresponding annihilation operator, 
$\hat c_{i\alpha}$. 
We can then use Wick's theorem to evaluate the expectation value and recast it in terms of a Grassmannian effective field theory 
with action:
\begin{equation}
 {\cal S} =  - \frac{1}{2}\sum_{ij} \bar \Psi_i (D^{-1}_0)_{ij} \Psi_j -u \sum_i t_i \;.
\label{eq:S_eff}
\end{equation}
Here $t_i = \prod_{\alpha} \bar c_{i\alpha} c_{i\alpha}$, with 
 $c_{i\alpha}$  a Grassman field, and $\bar \Psi_i = (\bar c_{i\alpha},c_{i\beta})$
denotes a ``Nambu spinor''. The 
expectation value of the particle density is given by 
$
\langle \hat c_{i\alpha}^+ \hat c_{i\alpha}  \rangle_G 
=  \langle n_{i\alpha} \rangle_{\cal S} -u\langle t_{i} \rangle_{\cal S}
$,
while the average of terms occurring in the kinetic and interaction energy are  given by  
$
\sum_{\alpha} \langle \hat c_{i\alpha}^+ \hat c_{j\alpha}  \rangle_G 
=  \sum_{\alpha} \langle \bar c_{i\alpha} c_{j\alpha} [1-(1-g)(d_{i\alpha} + d_{j\alpha}) +
 (1-g)^2 (d_{i\alpha} d_{j\alpha} +  \delta_{ij} d_{i\alpha})] \rangle_{\cal S}$,
and 
$\langle \hat n_{i\alpha} \hat n_{i\beta}  \rangle_G = 
 \langle n_{i\alpha} n_{i\beta} \rangle_{\cal S} -u \langle t_{i} \rangle_{\cal S}$, 
with  $n_{i\alpha}= \bar c_{i\alpha} c_{i\alpha}$ and  $d_{i\alpha} = \prod_{\beta\ne\alpha }n_{i\beta}$.
The Gutzwiller projector manifests in Eq.~(\ref{eq:S_eff}) as
 an effective interaction,  $u=1-g^2$, which is attractive for trionic correlations ($g>1$).
The propagator $D_0$ is given by 
\be
(D^{-1}_0)_{ij} = \left( \begin{array}{cc} G_0 & F_0 \\ F_0^{+} & -G_0^{+}\end{array}
  \right)_{ij}^{-1}\;,
\ee
with $G_{0 \; ij}^{\alpha \beta} \equiv -\langle \hat c_{i\alpha} \hat c_{j \beta}^+ \rangle_{BCS}$
and $F_{0 \; ij}^{\alpha \beta} \equiv  -\langle \hat c_{i\alpha} \hat c_{j \beta} \rangle_{BCS}$.

Similar to \cite{DMFT}, we can derive  identities 
that relate all expectation values above 
to the full  Green's function, $D_{ij}\equiv 
- \langle \Psi_i \bar \Psi_j \rangle_{\cal S}$  of the effective field theory. 
While we cannot  compute $D_{ij}$ analytically  in finite dimensions, 
a great simplification occurs in the $d\to\infty$ limit with $t\equiv t^*/\sqrt{d}$  
 (and $t^*=$ fixed), where the 
self energy becomes completely \emph{ local}~\cite{DMFT}.
In this limit, we can obtain  $D_{ij}$ and thus compute $E(\Delta,g,n,n_3)$
exactly, by deriving self-consistent integral
equations using an approach similar to   dynamical mean field theory \cite{DMFT,unpublished}. 

\begin{figure}
\begin{center}
\includegraphics[width=7.2cm]{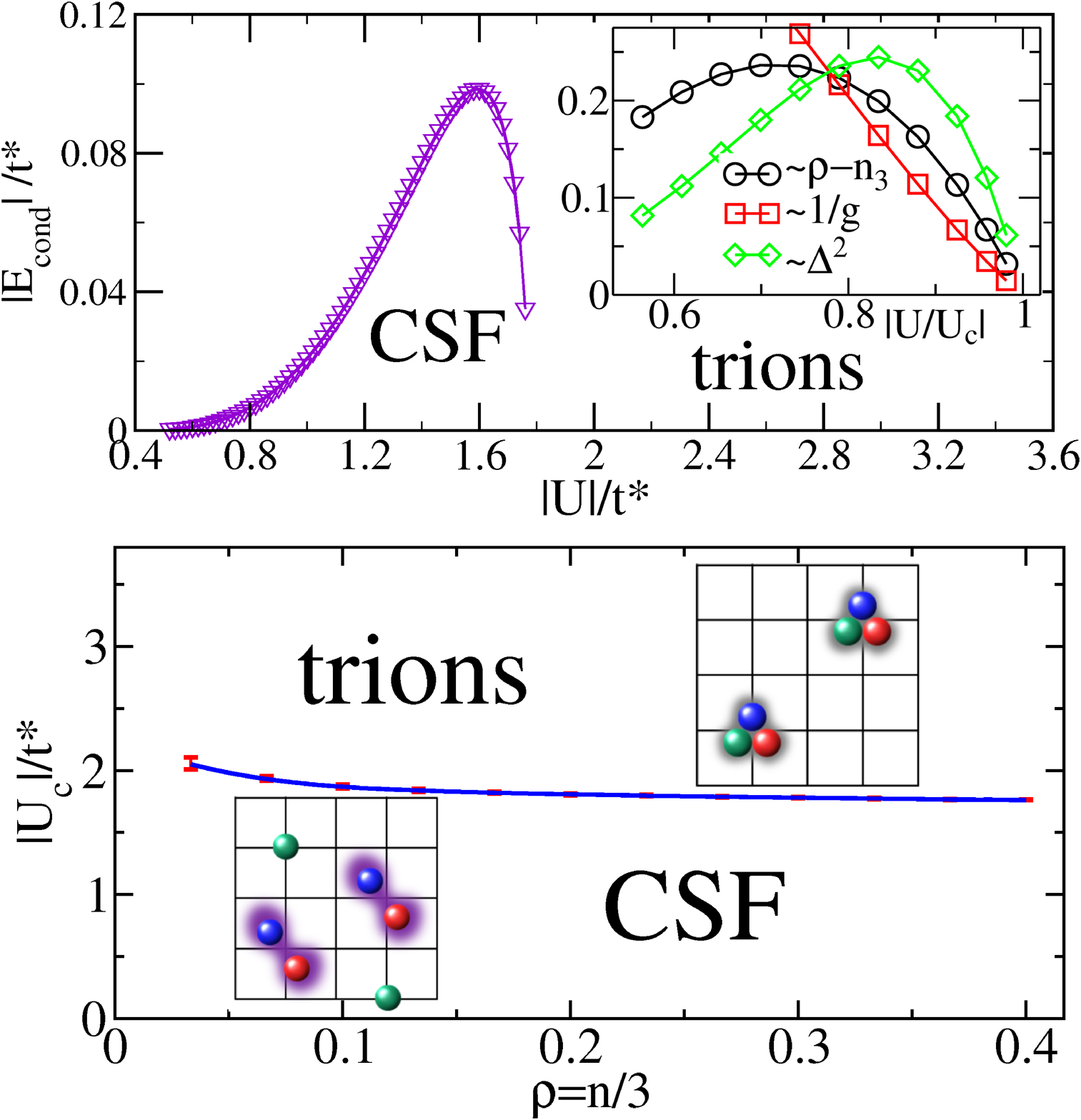}
\end{center}
\vspace*{-10pt}
\caption{
(color online). Upper panel: Condensation energy as a function of $|U|/t^*$ for filling $\varrho\equiv n/3=1/3$.
Note that $t^* = t \sqrt{d}$ is the typical kinetic energy of the fermions and determines the Fermi temperature $T_F$. 
The inset shows how the superconducing order parameter, $\Delta$, the deviation from equal
densities,  $\delta n_3 \equiv \varrho-n_3$, and $1/g$ scale to zero as one approaches the 
critical value $U_C/t^*\approx -1.774$. Lower panel: $U_C/t^*$ as a function of the filling factor $\varrho=n/3$. The phase diagram is  symmetric to $\varrho = 0.5$.
\label{fig:fixed_n}
}
\vspace*{-10pt}
\end{figure}

The upper panel in Fig.~\ref{fig:fixed_n} summarizes  our results for 
the filling $\varrho\equiv n/3=1/3$. 
We show the condensation energy per lattice site, defined as the energy difference 
between the superfluid state and the minimum energy state for the same filling 
and $\Delta=0$. 
For $|U| < |U_C|\approx -1.774 \, t^* $, the minimum occurs at finite values of $\Delta$ and $g>1$; thus, we find a CSF with built-in three-body correlations. 
As expected, the energy of the CSF has a minimum 
with $n_1 = n_2>n_3$. 
In other words, the 'spin operators'
defined in terms of the the Gell-Mann matrices as
$ {\hat c}^\dagger_{i} \lambda^\mu  {\hat c}_{i}$,   
tend to acquire a non-zero value, and the CSF
becomes also ``ferromagnetic''. With an equal number of 
atoms of each color, this implies phase separation; that is, superfluid \emph{ domains} will 
form with the order parameter in each domain ``pointing'' in a different direction \cite{Demler_work}. 
This tendency to form 
domains can also be captured through a Ginzburg-Landau analysis, to be discussed in a future publication \cite{unpublished}. 
Note that ferromagnetism appears only as a secondary order parameter, since the transition is driven by 
the local superfluid correlations. 

The transition temperature $T_C$ associated with the formation of the CSF is related to the condensation energy, $E_{\rm cond}$. 
As shown in Fig.~\ref{fig:fixed_n}, $E_{\rm cond}$ initially increases with 
$|U|$; however, it goes to zero  as one approaches the critical value, 
$U\to U_C$. At this point, the order parameter $\Delta$ also scales continuously to 0; 
thus, the point $U=U_C$ is a \emph{quantum critical point}.
At $U\to U_C$, the variational parameter $g$ diverges, and 
$|G\rangle$ reduces to a linear combination of states having 
trions (or no particles at all) at each lattice point. This $g\to\infty$ state
provides the  solution of minimum energy for the whole region $|U|> |U_C|$.
Note also that in the high-dimensional limit, the effective hopping amplitude of the trions 
vanishes as $\sim 1/d^{3/2}$. Therefore, trions are immobile in $d=\infty$ dimensions, and  
the energy of the trionic  state is simply $E_{\rm trion}/L  = n U = 3 \varrho U$, with 
$L$ the number of lattice sites. 
In $d=\infty$, the trionic state has also a finite residual entropy.
This entropy is, however, immediately removed once one considers a finite dimensional system where
trions can hop with some effective hopping $t_{\rm eff}\ne0$: Then trions are expected to 
form a Fermi liquid  with a Fermi temperature $T^*$. We expect the Fermi temperature 
to scale to 0 in finite dimensional lattices  as one approaches 
$U_C$ from the trionic side.  As shown in the inset of Fig.~\ref{fig:fixed_n}, the deviation from 
equal densities (``ferromagnetic polarization''), $\delta n_3 \equiv \varrho-n_3$,  also  vanishes as $U\to U_C$, 
since the condensation energy that  drives the 
accumulation of particles in the superfluid channels goes to zero at $U_C$.

This picture might, in principle, be modified through the intrusion of other phases. However, except for the half-filled fully-nested case not discussed here, there are no indications for this. At weak coupling, functional renormalization group \cite{Honerkamp1} does not detect any other instabilities.
The trionic repulsive Fermi liquid is apparently also stable, except possibly at $\rho \approx 1/2$, where charge ordering will take place.
We therefore believe that our variational calculation captures the basic structure of the phase diagram, which, however, might become decorated  by subdominant and possibly undetectable instabilities at lower energy scales.

At $T>0$, thermal fluctuations destroy the CSF order above a critical temperature. In 3-dimensional lattices, $T_C$ is expected to be finite and related to the condensation energy $E_{\rm cond}$. For $d<3$ dimensions, phase fluctuations completely destroy the  long-range superconducting order at $T>0$.

The phase diagram in Fig.~\ref{fig:fixed_n} parallels the famous
finite temperature phase diagram of QCD where a superconducting state occurs at 
large quark densities, corresponding to large kinetic energies in our case, while 
baryonic matter emerges at low densities, i.e., small kinetic energies \cite{QCDphase}. 
In QCD, however, the phase transition is believed to be of first order due to the long-range interaction 
generated by gluons. Furthermore, unlike quarks, the 3-color ultracold fermions considered here (e.g., $^6$Li)  
have no additional flavors and spin degrees of freedom. The CSF
emerging in our case has a non-trivial SU(3) color content and is analogous to the 
color superconducting phase in 2-flavor QCD where only two flavors of light quarks 
are considered \cite{2quark}.  
In the alternative theoretical scenario of 3-flavor QCD the superconducting state 
is expected to be color-flavor locked \cite{color-flavor}. 

The variational analysis can be performed for all densities. 
We summarize our results in the $T=0$ phase diagram in the lower panel in Fig.~\ref{fig:fixed_n}. 
Because of particle-hole symmetry the phase diagram is symmetric with respect to half-filling, $\varrho=1/2$. Close to half filling, Fermi surface nesting  could lead to additional phases not captured by our analysis. 

Experimentally, the critical temperature could be reached by implementing adiabatic cooling 
in the optical lattice \cite{HTC,Werner}. To leading order, this procedure leaves the ratio $T/T_F$ invariant as absolute temperature scales decrease. 
From Fig.~\ref{fig:fixed_n} we expect the CSF critical temperature to be 
$T_C/T_F \sim 0.1$ for the 3-color degenerate Fermi gas which has to be reached before the lattice is switched on. 
Realistic values for the dimensionless interaction strength are in the range $|U|/t \sim 1 - 100$. 
The full phase diagram can therefore be experimentally probed. 
Trion formation is expected to set in at temperatures of order $|U| \sim E_R$ where $E_R \approx 0.7 \mu K$ 
is the recoil energy of $^6$Li in an optical lattice \cite{bec-bcs-lattice}. 

Maybe the easiest way to detect the CSF 
is to break the SU(3) symmetry to U(1) and detect vortices and the superconducting
condensate fraction directly by pushing the system
through a Feshbach resonance \cite{bcs-imbalance}.
Furthermore, it should be also possible to observe the phase separation discussed above: 
For equal initial densities of all three internal states, domains will form where one of the states 
has a lower density than the others. This imbalance can be easily detected by light absorption, 
as already demonstrated for the BEC-BCS transition in a Fermi gas with overall spin imbalance \cite{bcs-imbalance}. 
Note that the system discussed here has equal total populations of all three hyperfine states, 
and the domain formation together with a \emph{local} imbalance occurs spontaneously. 

The quantum phase transition could also be observed 
via Bragg scattering \cite{Bragg}. With this technique, 
the dynamic structure factor $S(k, \omega)$ can be measured which  
is a suitable quantity for detecting the superfluid ground state \cite{HTC}. 
In the CSF, several Goldstone modes arise due to the reduced SU(2) symmetry of the ground state, and
one of them (the Anderson-Bogoliubov mode) is visible 
in $S(k, \omega)$ \cite{Honerkamp1}. This mode will become soft close to the transition as the order parameter vanishes 
and disappears in the trionic phase. 

The characteristic excitation spectrum of the trionic phase can be probed by ``shaking'' the 
optical lattice, i.e., applying a periodic amplitude modulation.  
This technique has already been used to determine the excitation spectrum of a bosonic Mott insulator 
\cite{shaking-exp,shaking-theo}. 
Deep in the trionic phase we expect a dominant excitation at a characteristic frequency $\omega = 2 |U|$ 
corresponding to the breaking up of trions, which becomes broadened by Fermi liquid quasiparticle 
excitations as one approaches the transition into the superfluid.  

Since neither the CSF  nor the trionic phase depends on a particular
filling,  we expect only small modifications of our results due to the trapping potential 
or the ensuing inhomogeneous density.
Moreover, for $U\approx U_C$ the Cooper pair size $\xi$ will be 
small compared to the trap diameter $R$. Thus, the trapping potential plays a minor 
role as in the measurement of the dynamic structure factor 
for interacting bosons \cite{Bragg}. 

We thank D. Rischke, P. Zoller, and especially
I. Bloch, E. Demler and M. Zwierlein for discussions
and comments. We have been supported by 
Hungarian grants OTKA Nos. NF061726, T046267, and T046303, 
by the German Science Foundation (DFG) grant HO 2407/2-1, 
and the Alexander von Humboldt Foundation.

\end{document}